\documentclass[a4paper,12pt]{article}
\usepackage{cite}
\newcommand{\be}{\begin{equation}}
\newcommand{\ee}{\end{equation}}
\newcommand{\bea}{\begin{eqnarray}}
\newcommand{\nn}{\nonumber}
\newcommand{\eea}{\end{eqnarray}}

\begin{document}

\begin{titlepage}
\begin{flushright}
UB-ECM-PF-03/23\\
IPM/P-2003/058
\end{flushright}
\begin{centering}
\vspace{.3in}
{\Large{\bf Self-Gravitational Corrections\\ to the Cardy-Verlinde Formula of Ach\'ucarro-Ortiz Black Hole}}
\\

\vspace{.5in} {\bf Mohammad R. Setare$^{1}$ and Elias C.
Vagenas$^{2}$ }\\
\vspace{.3in} $^{1}$\,Physics Dept., Inst. for Studies in Theo. Physics and
Mathematics (IPM)\\
P.O. Box 19395-5531, Tehran, Iran\\
rezakord@ipm.ir\\
\vspace{0.4in}

$^{2}$\, Departament d'Estructura i Constituents de la Mat\`{e}ria\\
and\\ CER for Astrophysics, Particle Physics and Cosmology\\
Universitat de Barcelona\\
Av. Diagonal 647, E-08028 Barcelona\\
Spain\\
evagenas@ecm.ub.es\\
\end{centering}

\vspace{0.6in}
\begin{abstract}
Recently, it was shown that the entropy of the black hole horizon in the Ach\'ucarro-Ortiz spacetime can be
described by the Cardy-Verlinde formula. In this paper, we compute the self-gravitational corrections to
the Cardy-Verlinde formula of the two-dimensional Ach\'ucarro-Ortiz black hole. These corrections stem from the
effect of self-gravitation and they are derived in the context of Keski-Vakkuri, Kraus and Wilczek (KKW)
analysis. The black hole under study is therefore treated as a dynamical background. The self-gravitational
corrections to the entropy as given  by the Cardy-Verlinde formula
of Ach\'ucarro-Ortiz black hole, are found to be positive.
This result provides evidence in  support of the claim that the holographic
bound is not universal in the framework of two-dimensional gravity models.
\end{abstract}
\end{titlepage}

\newpage

\baselineskip=18pt
\section*{Introduction}
In 1992 Ba$\tilde{n}$ados, Teitelboim and Zanelli (BTZ)
\cite{banados1,banados2} showed that ($2+1$)-dimensional gravity
has a black hole solution. This black hole is described by two parameters,
its mass $M$ and its angular momentum (spin) $J$. It is locally
anti-de-Sitter space and thus it differs from Schwarzschild and
Kerr solutions in that it is asymptotically anti-de-Sitter instead
of flat spacetime. Additionally, it has no curvature singularity at the
origin. AdS black holes, which are members of the two-parametric family
of BTZ black holes, play a central role in AdS/CFT conjecture
\cite{maldacena1} and also in brane-world scenarios
\cite{RS1,RS2}. Specifically AdS(2) black hole is most interesting in the context
of string theory and black hole physics \cite{sen,strominger1,strominger2}.
\par
Concerning the quantum process called Hawking effect \cite{hawking1} much work has been done using a fixed
background during the emission process. The idea of Keski-Vakkuri, Kraus and Wilczek (KKW)
\cite{KKW1}-\hspace{-0.1ex}\cite{KKW4} is to view the black hole background as dynamical by treating the Hawking radiation as
a tunnelling process. The energy conservation is the key to this description. The total (ADM) mass is kept fixed
while the mass of the black hole under consideration decreases due to the emitted radiation. The effect of
this modification gives rise to additional terms in the formulae concerning the known results for
black holes \cite{correction1}-\hspace{-0.1ex}\cite{elias5}; a nonthermal partner to the
thermal spectrum of the Hawking radiation shows up.
\par
Holography is believed to be one of the fundamental principles of the true
quantum theory of gravity \cite{{HOL},{RAP}}. An explicitly
calculable example of holography is the much--studied anti-de
Sitter (AdS)/Conformal Field Theory (CFT) correspondence. More
recently, it has been proposed in a manner analogous
with the AdS$_{d}$/CFT$_{d-1}$ correspondence, that quantum gravity in a
de Sitter (dS) space is dual to a certain
Euclidean  CFT living on a spacelike boundary of the
dS space~\cite{Strom} (see also earlier works
\cite{mu2}-\hspace{-0.1ex}\cite{Bala}). Following this proposal, some
investigations on the dS space have been carried out recently
\cite{Mazu}-\hspace{-0.1ex}\cite{set3}.
\par
The Cardy-Verlinde formula recently proposed by  Verlinde \cite{Verl}, relates the entropy of a  certain CFT to
its total energy and Casimir energy in arbitrary dimensions. In the spirit of AdS$_{d}$/CFT$_{d-1}$ and
dS$_{d}$/CFT$_{d-1}$ correspondences, this formula has been shown to hold exactly for the cases of topological dS,
Schwarzschild-dS, Reissner-Nordstr\"om-dS, Kerr-dS and Kerr-Newman-dS black holes.
\par
Recently, much interest has been taken  in computing the quantum corrections to the
Bekenstein-Hawking entropy $S_{BH}$ \cite{med,muk,lid}. In a recent work Carlip \cite{carlip} has
deduced the leading order quantum correction to the classical Cardy formula. The Cardy formula
follows from a saddle-point approximation of the partition function for a two-dimensional CFT.
This leads to the theory's density of states which is related to the partition function
by way of a Fourier transform.  In \cite{med}, Medved has employed Carlip's formulation
to the case of a generic model of two-dimensional gravity with coupling to a dilaton field.
In the present paper we study the semi-classical gravitational corrections to the Cardy-Verlinde
formula due to the effect of self-gravitation.
\par
The remainder of this paper is organized as follows. In Section 1, we make a short review of the two-dimensional
Ach\'ucarro-Ortiz black hole. We present for the afore-mentioned black hole, expressions for its mass, angular
momentum, angular velocity, temperature, area and entropy. In Section 2, we compute the self-gravitational
corrections to the entropy of the two-dimensional Ach\'ucarro-Ortiz black hole which is described by the
Cardy-Verlinde formula. Finally, in Section 3  we briefly summarize our results and some concluding remarks are
made.
\section{Ach\'ucarro-Ortiz  Black Hole}
 The black hole
solutions of Ba\~nados, Teitelboim and Zanelli \cite{banados1,banados2} in $(2+1)$ spacetime dimensions are
derived from a three dimensional theory of gravity \be S=\int dx^{3} \sqrt{-g}\,({}^{{\small(3)}} R+2\Lambda) \ee
with a negative cosmological constant ($\Lambda=1/l^2 >0$).
\par\noindent
The corresponding line element is \be ds^2 =-\left(-M+ \frac{r^2}{l^2} +\frac{J^2}{4 r^2} \right)dt^2
+\frac{dr^2}{\left(-M+ \displaystyle{\frac{r^2}{l^2} +\frac{J^2}{4 r^2}} \right)} +r^2\left(d\theta
-\frac{J}{2r^2}dt\right)^2 \label{metric}\ee It is obvious that there are many  ways to reduce the three
dimensional BTZ black hole solutions to the two dimensional charged and uncharged dilatonic black holes
\cite{ortiz,lowe}. The Kaluza-Klein reduction of the $(2+1)$-dimensional metric (\ref{metric}) yields a
two-dimensional line element:
 \be ds^2 =-g(r)dt^2 +g(r)^{-1}dr^2
\label{metric1}\ee where \be g(r)=\left(-M+\frac{r^2}{l^2}
+\frac{J^2}{4 r^2}\right)\label{metric2}
 \ee
with $M$ the mass of the two-dimensional Ach\'ucarro-Ortiz black hole, $J$ the angular momentum (spin)
 of the afore-mentioned black hole and $-\infty<t<+\infty$, $0\leq r<+\infty$, $0\leq \theta <2\pi$.
\par \noindent
The outer and inner horizons, i.e. $r_{+}$ (henceforth simply
black hole horizon) and $r_{-}$ respectively, concerning the
positive mass black hole spectrum with spin ($J\neq 0$) of the
line element (\ref{metric}) are given as  \be
r^{2}_{\pm}=\frac{l^2}{2}\left(M\pm\sqrt{M^2 -
\displaystyle{\frac{J^2}{l^2}} }\right) \label{horizon1} \ee and
therefore, in terms of the inner and outer horizons, the black
hole mass and the angular momentum are given, respectively, by
\be
M=\frac{r^{2}_{+}}{l^{2}}+\frac{J^{2}}{4r^{2}_{+}}
\label{mass}
\ee
and
\be
J=\frac{2\, r_{+}r_{-}}{l}
\label{ang}
\ee
with the
corresponding angular velocity to be
\be
\Omega=\frac{J}{2
r^{2}}\label{angvel}\hspace{1ex}.
\ee
\par\noindent
The Hawking temperature $T_H$ of the black hole
horizon is \cite{kumar1}
\bea
T_H &=&\frac{1}{2\pi
r_{+}}\sqrt{\left(\displaystyle{\frac{
r_{+}^2}{l^2}+\frac{J^2}{4r_{+}^2}}\right)^2-\displaystyle{\frac{J^2}{l^2}}}\nn\\
&=&\frac{1}{2\pi r_{+}}\left(\displaystyle{\frac{ r_{+}^2}{l^2}-\frac{J^2}{4r_{+}^2}}\right)
\label{temp1}
\hspace{1ex}.
\eea
\par\noindent
The area $\mathcal{A}_H$ of the black hole horizon is
\bea
\mathcal{A}_{H}&=&\sqrt{2}\,\pi l
\left(M+\sqrt{M^2 -\displaystyle{\frac{J^2}{l^2}} }\right)^{1/2}\label{area1}\\
&=& 2\pi r_{+}
\label{area2}
\eea and thus the entropy of the two-dimensional Ach\'ucarro-Ortiz black hole, if we
employ the well-known Bekenstein-Hawking area formula ($S_{BH}$) for the entropy
\cite{bekenstein1,bekenstein2,hawking3}, is given by
\be
S_{BH}=\frac{1}{4\hbar G} \mathcal{A}_H
\hspace{1ex}.
\label{entropy1}
\ee
Using the BTZ units where $8\hbar G =1 $, the entropy of the two-dimensional
Ach\'ucarro-Ortiz black hole  takes the form
\be
S_{BH}=4 \pi r_{+}
\label{entr1}
\hspace{1ex}.
\ee
\section{Self-gravitational corrections to Cardy-Verlinde formula}
In this section we compute the self-gravitational corrections to the entropy of the two-dimensional
Ach\'ucarro-Ortiz black hole (\ref{entr1}) described by the Cardy-Verlinde
formula
\be
S_{CFT}=\frac{2\pi
R}{\sqrt{ab}}\sqrt{E_{C}\left(2E-E_{C}\right)}
\label{cvf}\hspace{1ex}.
\ee
The total energy $E$ may be written as the sum of two terms
\be
E(S, V)=E_{E}(S, V)+\frac{1}{2}E_{C}(S,V)
\label{ext}
\ee
where $E_{E}$ is the purely extensive part of the total energy $E$ and $E_{C}$ is the Casimir energy.
\par\noindent
The Casimir energy is derived by the violation of the Euler relation
\be
E_{C}=2 E-T_{H}S_{BH}-\Omega_{+} J
\label{euler} \ee which now  will be modified due to the self-gravitation effect as \be E_{C}=2
E-T_{bh}S_{bh}-\Omega_{bh}J_{bh}
\label{euler1}
\hspace{1ex}.
\ee
In the context of KKW analysis, it is easily seen
that\footnote{For the explicit computation of the modified thermodynamical quantities
see \cite{correction5}}
\bea
T_{bh}S_{bh}&=&T_{H}S_{BH}\left(1-\omega\frac{M}{2\sqrt{M^2 -\displaystyle{\frac{J^2}{l^2}}}\,
\left(M+\sqrt{M^2 -\displaystyle{\frac{J^2}{l^2}}}\right)}\right)\\
&=&T_{H}S_{BH}\left(1-\omega\frac{Ml^{4}}{4r_{+}^{2}\left(r_{+}^{2}-r_{-}^{2}\right)}\right)
\eea
where $\omega$ is the emitted shell of energy radiated outwards the black hole horizon.
\par\noindent
Thus, the second term in (\ref{euler1}) can be written as
\be
T_{bh}S_{bh}=T_{H}S_{BH}-\omega \frac{M}{\left(M+\sqrt{M^2 -\displaystyle{\frac{J^2}{l^2}}}\right)}
\label{corts}
\ee
where
\be
T_{H}S_{BH}=2\left(\displaystyle{\frac{
r_{+}^2}{l^2}-\frac{J^2}{4r_{+}^2}}\right)
\hspace{1ex}.
\label{ts}
\ee
Additionally, one can easily check that in the context of KKW analysis
the modified angular momentum (computed up to first order in $\omega$) is given as
\be
J_{bh}=J\left(1-\epsilon_{1} \omega\right)
\label{corang}
\ee
where $\epsilon_{1}$ is a parameter\footnote{This parameter is small
for sufficiently large mass of the two-dimensional Ach\'ucarro-Ortiz
black hole (as it is expected for such a
 semiclassical analysis here employed since the radiating
matter is viewed as point particles).}
\bea
\epsilon_{1}& =&
\frac{1}{2\sqrt{M^2 -\displaystyle{\frac{J^2}{l^2}}}}\\
&=&\frac{l^2}{2\left(r_{+}^{2}-r_{-}^{2}\right)}\hspace{1ex}.
\eea
The modified angular velocity on the black hole horizon
(computed also up to first order in $\omega$) is
\be
\Omega_{bh}=\frac{\Omega_{+}}{\left(1-\epsilon_{1} \omega\right)}
\label{corangvel}
\ee
where $\Omega_{+}$ is the angular velocity evaluated on the black hole horizon
\be
\Omega_{+}=\frac{J}{2
r^{2}_{+}}\hspace{1ex}.
\ee
Therefore, the third term in (\ref{euler1}) is given as
\be
\Omega_{bh}J_{bh} =\Omega_{+}J
\label{corjo}
\ee
where
\be
\Omega_{+}J=\frac{J^2}{2r^{2}_{+}}
\label{jo}
\hspace{1ex}.
\ee
At this point it is necessary to stress that we shall consider no self-gravitational
 corrections to the total energy $E$ of the system under study as well as to the radius
 which takes the form \cite{e-m}
\be
R= 2 r_{+}\left(\frac{l}{J}\right)\sqrt{ab}
\label{radius3}
\hspace{1ex}.
\ee
The Casimir energy, substituting (\ref{corts}) and (\ref{corjo}) in  (\ref{euler1}), is given as
\bea
E_{C}&=&\frac{J^{2}}{2r_{+}^2}+\omega \frac{M}{\left(M+\sqrt{M^2 -\displaystyle{\frac{J^2}{l^2}}}\right)}\\
&=&\frac{J^{2}}{2r_{+}^2}+2\epsilon_{2} M\,\omega
\label{euler2}
\eea
where $\epsilon_{2}$ is a parameter given
by\footnote{This is also a small parameter for sufficiently large mass of
the two-dimensional Ach\'ucarro-Ortiz black hole.}
\bea
\epsilon_{2}& =&
\frac{1}{2\left(M+\sqrt{M^2 -\displaystyle{\frac{J^2}{l^2}}}\right)}\\
&=&\frac{l^2}{4 r_{+}^{2}}
\hspace{1ex}.
\eea

Additionally, it is evident that the quantity
$2E-E_{C}$ is given, by substituting again equations (\ref{corts}) and (\ref{corjo}) in (\ref{euler1}), as
\bea
2E-E_{C}&=&
2\frac{r_{+}^2}{l^2}-\omega \frac{M}{\left(M+\sqrt{M^2 -\displaystyle{\frac{J^2}{l^2}}}\right)}\\
&=&2\frac{r_{+}^2}{l^2}-2\epsilon_{2} M\,\omega
\label{core-ec}
\hspace{1ex}.
\eea
Apart from the Casimir energy, the purely extensive part of the total energy $E_{E}$ will also be modified
due to the effect of self-gravitation. Thus, it takes the form
\bea
E_{E}&=& \frac{r_{+}^2}{l^2}-\omega \frac{M}{2\left(M+\sqrt{M^2 -\displaystyle{\frac{J^2}{l^2}}}\right)}\\
&=&\frac{r_{+}^2}{l^2}-\epsilon_{2} M\,\omega
\label{corexten2}
\eea
whilst it can also be written as \cite{e-m}
\bea
E_{E}&=&\frac{a}{4\pi R}S_{bh}^{2}\\
&=&\frac{4\pi a}{R}r^{2}_{out}\\
&=&\frac{4\pi a}{R}r^{2}_{+}(1-2\epsilon_{1}\omega)
\label{corexten3}
\hspace{1ex}.
\eea
\par\noindent
We substitute  expressions (\ref{radius3}), (\ref{euler2}) and (\ref{core-ec})
which were computed to first order in $\omega $ (in the framework of KKW analysis)
in the Cardy-Verlinde formula in order that self-gravitational corrections to be considered,
\be
S_{CFT}=\frac{2\pi}{\sqrt{ab}}2 r_{+}
\left(\frac{l}{J}\sqrt{ab}\right)
\sqrt{\left(\frac{J^2}{2r^{2}_{+}}+2\epsilon_{2}M\,\omega\right)
\left(\frac{2r^{2}_{+}}{l^2}-2\epsilon_{2}M\,\omega\right)}
\ee
and consequently, the first-order self-gravitationally corrected
Cardy-Verlinde formula of the two-dimensional Ach\'ucarro-Ortiz black hole takes the form
\be
S_{CFT}=S_{BH}\sqrt{1+\epsilon_{3}\omega}
\hspace{1ex}.
\ee
where
\bea
\epsilon_{3}&=&4\frac{M l^{2}}{J^{2}}
\left(\frac{r_{+}^{2}}{l^2}-\frac{J^2}{4r^{2}_{+}}\right)\epsilon_{2}\\
&=&2M\left(\frac{2\pi l^{2}}{J}\right)^{2}\frac{T_{H}}{S_{BH}}
\eea
It is easily seen that the parameter $\epsilon_{3}$ is positive and therefore
the self-gravitational corrections are also positive.
\par\noindent
It should be pointed out that in the context of KKW analysis the self-gravitational corrections
to the entropy as described by the Cardy-Verlinde formula ($S_{CFT}$) of the
two-dimensional Ach\'ucarro-Ortiz black hole are  different from the ones to
the corresponding Bekenstein-Hawking entropy ($S_{BH}$).
This is expected since in order to evaluate the corrections to the entropy as described
by the Cardy-Verlinde formula ($S_{CFT}$), we have taken into account not only corrections to the
Bekenstein-Hawking entropy but also to all quantities appearing in the Cardy-Verlinde formula,
i.e. the Hawking temperature ($T_{H}$), the angular momentum ($J$) and the corresponding
angular velocity ($\Omega_{+}$).
Furthermore, the entropy of the two-dimensional Ach\'ucarro-Ortiz black hole ($S_{CFT}$)
described in the context of KKW analysis by the semiclassically corrected
Cardy-Verlinde formula violates the holographic bound \cite{HOL}, i.e.
\be
S_{CFT}>S_{BH}>S_{bh}
\hspace{1ex}.
\ee
\section{Conclusions}
In this work we have evaluated the semiclassical corrections to the entropy of two-dimensional
Ach\'ucarro-Ortiz black hole as described by the Cardy-Verlinde formula. These
corrections are due to the self-gravitation effect. They are derived in the context of
KKW analysis and we have kept up to linear terms in the energy of the emitted
massless particle. The afore-mentioned gravitational background is treated as a
dynamical one and the self-gravitational corrections to its entropy are found to be positive.
This result is a direct violation to the holographic bound.
\par\noindent
A couple of comments are in order concerning this violation.
Firstly, it is known that the entropy of BTZ black hole does not violate the entropy bounds.
Thus, it is expected that the Ach\'ucarro-Ortiz black hole which is derived by a
dimensional reduction of the BTZ black hole and shares the same
thermodynamical formulas with BTZ black hole, will also respect the entropy bounds.
On the contrary, in the context of KKW analysis it was
proven\footnote{Compare results in \cite{correction5} with the corresponding ones in \cite{correction6}.}
that the modified entropy of  Ach\'ucarro-Ortiz black hole is different even to first  order from
the corresponding modified entropy of BTZ black hole.
Secondly, Mignemi recently claimed \cite{mignemi} that the existence of a holographic bound
depends on the dynamics of the specific model of gravity
in contrast to the Bekenstein bound which is inherent to the definition
of black hole thermodynamics in any metric theory of gravity.
Therefore, our result provides evidence in support of the claim
that the holographic bound is not universal for 2D gravity models.
\section{Acknowledgments}
The work of E.C.V. has been supported by the
European Research and Training Network ``EUROGRID-Discrete Random
Geometries: from Solid State Physics to Quantum Gravity"
(HPRN-CT-1999-00161).

\end{document}